\documentclass[11pt,draftcls,onecolumn]{IEEEtran}
\usepackage{amsmath,graphicx,amsfonts,amsthm,textcomp,cite}
\usepackage{amssymb}
\usepackage{color}
\usepackage{comment}
\usepackage{subcaption}
\usepackage{url}
\usepackage{epstopdf}

\newtheoremstyle{mythmstyle}
{3pt}
{3pt}
{\itshape}
{}
{\bfseries}
{:}
{.5em}
{}
\theoremstyle{mythmstyle}
\newtheorem{mylemma}{Lemma}
\newtheorem{mythm}{Theorem}

\newtheoremstyle{mythmstyle2}
{3pt}
{3pt}
{}
{}
{\bfseries}
{:}
{.5em}
{}
\theoremstyle{mythmstyle2}
\newtheorem{myex}{Example}
\newtheorem{mysex}{Numerical Example}

\begin{document}

\title{Performance Limits of Segmented Compressive Sampling:
Correlated Samples versus Bits}

\author{Hao~Fang, 
Sergiy~A.~Vorobyov, 
and Hai~Jiang
\thanks{H.~Fang is with the Department of Electrical Engineering, University of
Washington, Seattle, WA 98195, USA;
e-mail: {\tt hfang@uw.edu}.
S.~A.~Vorobyov is with the Department of Signal Processing and Acoustics, Aalto
University, Espoo, FI-02150, Finland;
email: {\tt svor@ieee.org}.
H.~Jiang is with the Department of Electrical and Computer Engineering,
University of Alberta, Edmonton, AB T6G 2V4, Canada; 
e-mail: {\tt hai1@ualberta.ca}. S.~A.~Vorobyov is the corresponding author.}%
}
\maketitle

\begin{abstract}
This paper gives performance limits of the segmented compressive sampling 
(CS) which collects correlated samples. It is shown that the effect 
of correlation among samples for the segmented CS can be characterized by 
a penalty term in the corresponding bounds on the sampling rate. Moreover, 
this penalty term is vanishing as the signal dimension increases. It means that 
the performance degradation due to the fixed correlation among samples 
obtained by the segmented CS (as compared to the standard CS with equivalent size 
sampling matrix) is negligible for a high-dimensional signal. In combination with 
the fact that the signal reconstruction quality improves with additional 
samples obtained by the segmented CS (as compared to the standard CS with 
sampling matrix of the size given by the number of original uncorrelated 
samples), the fact that the additional correlated samples also provide 
new information about a signal is a strong argument for the segmented CS.
\end{abstract}


\section{Introduction}
\label{sec:intro}
The theory of compressive sampling/sensing (CS) concerns of the possibility 
to recover a signal ${\bf x} \in \mathbb{R}^n$ from $m$ ($\ll n$) noisy 
samples
\begin{equation}
  {\bf y} = {\bf \Phi x}+{\bf z} 
  \label{eq:cs_sys}
\end{equation}
where ${\bf y} \in \mathbb{R}^m$ is the sample vector, ${\bf
\Phi} \in \mathbb{R}^{m \times n}$ is the sampling matrix, and
${\bf z} \in \mathbb{R}^m$ is the random noise
vector~\cite{Candes2005a,Candes2006c,Candes2008}.
In a variety of settings, the signal ${\bf x}$ is an $s$-sparse signal, i.e.,
only $s$ ($\ll n$) elements in the signal are nonzero; 
in some other settings, the signal ${\bf x}$ is sparse in some orthonormal basis
${\bf \Psi}$, i.e., the projection of ${\bf x}$ onto ${\bf \Psi}$ is an
$s$-sparse signal.
An implication of the CS theory is that an analog signal (not necessarily
band-limited) can be recovered from fewer samples than that required by the
Shannon's sampling theorem, as long as the signal is sparse in some orthonormal
basis~\cite{Candes2005a,Candes2006c,Candes2008,Donoho2006}.
This implication gives birth to the analog-to-information conversion
(AIC)~\cite{Laska2007,Candes2008}.
The AIC device consists of several parallel branches of mixer and integrators
(BMIs) performing random modulation and pre-integration (RMPI).
Each BMI measures the analog signal against a unique random sampling waveform
by multiplying the signal to the sampling waveform and then integrating the
result over the sampling period $T$.
Essentially, each BMI acts as a row in the sampling matrix ${\bf \Phi}$, and
the collected samples correspond to the sample vector ${\bf y}$ in
(\ref{eq:cs_sys}).
Therefore, the number of samples that can be collected by the traditional
BMI-based AIC device is equal to the number of available BMIs. The RMPI-based 
design has already led to first working hardware devices for AIC, see for 
example \cite{Becker}. Regarding the important areas within CS, it is worth 
quoting Becker's thesis \cite{Becker}: ``The real significance of CS was a 
change in the very manner of thinking ... Instead of viewing $\ell_1$ 
minimization as a post-processing technique to achieve better signals, CS has 
inspired devices, such as the RMPI system ..., that acquire signals in a 
fundamentally novel fashion, regardless of whether $\ell_1$ minimization is 
involved.'' However, in the case of noisy samples it is always beneficial 
to have more samples for better signal reconstruction.

Recently, Taheri and Vorobyov developed a new AIC structure using the segmented CS
method to collect more samples than the number of
BMIs~\cite{Taheri2010, Taheri2011}.
In the segmented CS-based AIC structure, the integration period $T$ is divided
into $t$ sub-periods, and sub-samples are collected at the end of each
sub-period.
Each BMI can produce a sample by accumulating $t$ sub-samples within
the BMI. 
Additional samples are formed by accumulating $t$ sub-samples from
different BMIs at different sub-periods.
In this way, more samples than the number of BMIs can be obtained.
The additional samples can be viewed as obtained from an extended
sampling matrix whose rows consist of permuted segments of the original
sampling matrix \cite{Taheri2011}.
Clearly, the additional samples are correlated with the original
samples and possibly with other additional samples. 
A natural question is whether and how these additional samples can 
bring new information about the signal to enable a higher quality recovery.
This motivates us to analyze and quantify the performance limits of the 
segmented CS in this paper.

Various theoretical bounds have been obtained for the problems of sparse
support recovery.
In \cite{Wainwright2009, Wang2010a, Reeves2008, Akcakaya2010, Aeron2010},
sufficient and necessary conditions have been derived for exact support recovery
using an optimal decoder which is not necessarily computationally tractable.
The performance of a computationally tractable algorithm named
$\ell_1$-constrained quadratic programming has been analyzed in
\cite{Wainwright2009a}.
Partial support recovery has been analyzed in \cite{Reeves2008, Reeves2013,
Akcakaya2010}.
In \cite{Akcakaya2010}, the recovery of a large fraction of the signal energy 
has been also analyzed.

Meanwhile, sufficient conditions have been given for the CS recovery with
satisfactory distortion using convex programming \cite{Candes2006c,
Baraniuk2008, Haupt2006}.
By adopting results in information theory, sufficient and necessary conditions
have also been derived for CS, where the reconstruction algorithms are not
necessarily computationally tractable.
Rate-distortion analysis of CS has been given in \cite{Fletcher2007,
Akcakaya2010, Sarvotham2006}.
In \cite{Sarvotham2006}, it has been shown that when the samples are
statistically independent and all have the same variance, the CS 
system is optimal in terms of the required sampling rate in order to achieve
a given reconstruction error performance.
However, some CS systems, e.g., the segmented CS architecture in
\cite{Taheri2011}, have correlated samples. 

In \cite{Candes2010}, the performance of CS with coherent and redundant
dictionaries has been studied. Under such setup, the resulting samples can be 
correlated with each other due to the non-orthogonality and redundancy of the 
dictionary. Unlike the case studied in \cite{Candes2010}, the correlation between
samples in the segmented CS is caused by the extended sampling matrix
whose rows consist of permuted segments of the original sampling matrix
\cite{Taheri2011}. It has been 
shown in \cite{Taheri2011, TaheriAsilomar} that the additional correlated 
samples help to reduce the signal reconstruction mean-square error (MSE), 
where the study has been performed based on the empirical risk minimization 
method for signal recovery, for which the least absolute shrinkage and 
selection operator (LASSO) method, for example, can be viewed as one of the 
possible implementations \cite{Haupt2006}.
Considering the attractive features of the segmented CS architecture, it is
necessary to analyze its performance limits where there is a fixed correlation
among samples caused by the extended sampling matrix.

In this paper, we derive performance limits of the segmented CS where
the samples are correlated. It will be 
demonstrated that the segmented CS is not a post-processing on the samples
as post-processing cannot add new information about the signal. 
In our analysis, the interpretation of the sampling matrix as a channel 
will be employed to obtain the capacity and distortion rate expressions for
the segmented CS.
It will make it easily visible how the segmented CS brings more
information about the signal - essentially, by using an extended (although 
correlated) channel/sampling matrix. Moreover, it will be 
shown that the effect of correlation among samples can
be characterized by a penalty term in a lower bound on the sampling rate.
Such penalty term will be shown to vanish as the length of the signal $n$
goes to infinity, which means that the influence of the fixed correlation among
samples is negligible for a high-dimensional signal. With such 
result to establish, we aim to verify the advantage of the segmented CS architecture, 
since it requires fewer BMIs, while achieving almost the same performance as the
non-segmented CS architecture that has a much larger number of BMIs.
We also aim at showing that as the number of additional samples correlated with the
original samples increases, the required number of original uncorrelated
samples decreases while the same distortion level is achieved.

The remainder of the paper is organized as follows.
Section~\ref{sec:prob_form} describes the mathematical setting considered in the
paper and provides some preliminary results.
The main results of this paper are presented in Section~\ref{sec:main_results},
followed by the numerical results in Section~\ref{sec:num_results}.
Section~\ref{sec:concl} concludes the paper.
Lengthy proofs of some results are given in Appendices
after Section~\ref{sec:concl}.

\section{Problem formulation, assumptions and preliminaries}
\label{sec:prob_form}
\subsection{Preliminaries}
The CS system is given by (\ref{eq:cs_sys}).
We use an $m \times 1$ random vector ${\bf w}$ to denote the
noiseless sample vector, i.e., 
\begin{equation}
  {\bf w = \Phi x}.
  \label{eq:noiseless_msr}
\end{equation}

\begin{figure*}[ht]
  \centering
  \includegraphics[width=0.6\textwidth]{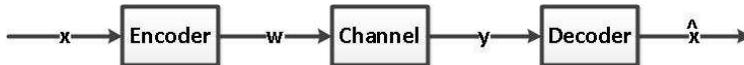}
  \caption{Block diagram of a CS system.}
  \label{fig:block_diag}
\end{figure*}

Thus, the signal ${\bf x}$, the noiseless sample vector ${\bf
w}$, the noisy sample vector ${\bf y}$ and the reconstructed
signal $\hat{\bf x}$ form a Markov chain, i.e., ${\bf x} \rightarrow
{\bf w} \rightarrow {\bf y} \rightarrow \hat{\bf x}$, as shown in
\figurename\,\ref{fig:block_diag}, where the CS
system is viewed as an information theoretic channel.

In this paper, we consider an additive white Gaussian noise channel,
i.e., the noise ${\bf z} \in \mathbb{R}^m$ consists of $m$
independent and identically distributed (i.i.d.)
$\mathcal{N}(0, 1)$ random variables.
Accordingly, the average per sample signal-to-noise ratio (SNR),
denoted as $\gamma$, can be defined as the ratio of the average
energy of the noiseless samples ${\bf w}$ to the average energy
of the noise ${\bf z}$, i.e.,
\begin{equation}
  \gamma \overset{\triangle}= 
  \frac{\mathbb{E}[||{\bf w}||^2_2]}{\mathbb{E}[||{\bf z}||_2^2]} 
  = \frac{\mathbb{E}[||{\bf w}||^2_2]}{m}
  \label{eq:SNR}
\end{equation}
where $\mathbb{E}[\cdot]$ denotes the expectation of a random variable
and $||~\cdot~||_2$ stands for the $\ell_2$-norm of a vector.

Assuming that all elements of ${\bf w}$ have the same
expected value $\mu_W$ and using the assumption that the signal
and noise are uncorrelated, the SNR can be written as
\begin{equation}
  \gamma = \frac{{\rm tr}({\bf \Sigma}_W) + m\mu_W^2}{m}
  \label{eq:SNR2} 
\end{equation}
where ${\bf \Sigma}_W$ denotes the covariance matrix of ${\bf w}$, and
${\rm tr}(\cdot)$ refers to the trace of a matrix.
So we have ${\rm tr}({\bf \Sigma}_W) = m\gamma -
m\mu_W^2$.
According to \cite{Sarvotham2006}, the channel capacity,
i.e., the number of bits per compressed sample that can be
transmitted reliably over the channel in the CS system, satisfies
\begin{equation}
  C \leq \frac{1}{2} \log (1+\gamma-\mu_W^2)\quad 
  {\rm bits/sample}.
  \label{ineq:capacity}
\end{equation}
Throughout this paper, the base of the logarithm is 2.
The equality in (\ref{ineq:capacity}) is achieved when ${\bf \Sigma}_W$ is diagonal and the
diagonal entries are all equal to $\gamma-\mu_W^2$.
In other words, the equality is achieved when the samples in
${\bf w}$ are statistically independent and have the same variance
equal to $\gamma-\mu_W^2$.
Based on this result, \cite{Sarvotham2006} gives a lower bound on the
sampling rate $\delta \overset{\triangle}= m/n$ when a
distortion $D$ is achievable, that is,
\begin{equation}
  \delta \geq \frac{2R(D)}{\log(1+\gamma-\mu_W^2)}
  \label{lb:msr_rate}
\end{equation}
as $n \rightarrow \infty$, where $R(D)$ is the
rate-distortion function, which gives the minimal number of bits per
source symbol needed in order to recover the source sequence within a
given distortion $D$, and $D \overset{\triangle}{=} \mathbb{E}[d({\bf x}, 
\hat{\bf x})]$ is the average distortion achieved by the CS system. 
Here the distortion between two $n \times 1$ vectors ${\bf x}$ and
$\hat{\bf x}$ is defined by 
\begin{equation}
  d({\bf x}, \hat{\bf x}) = \frac{1}{n} \sum_{i=1}^n d(x_i, \hat{x}_i)	
  \label{eq:dist_msr}
\end{equation}
where $x_i$ and $\hat{x}_i$ denote, respectively, the $i$-th elements
of ${\bf x}$ and $\hat{\bf x}$, and $d({\bf x}, \hat{\bf x})$
and $d(x_i, \hat{x}_i)$ are the distortion measure between two vectors
and two symbols, respectively.

However, when the samples in the noiseless sample vector ${\bf w}$ are 
correlated, i.e., ${\bf \Sigma}_W$ is not a diagonal matrix, the upper 
bound on the channel capacity $C$ in (\ref{ineq:capacity}), and 
accordingly the lower bound on the sampling rate $\delta$ in 
(\ref{lb:msr_rate}), can never be achieved.
In this paper, we aim at showing the effects of
sample correlation on these bounds. 

\subsection{Stochastic Signal Assumptions}
Consider the following assumptions on the random vector ${\bf x} \in
\mathcal{Q} \subseteq \mathbb{R}^n$ where $\mathcal{Q}$ is a compact
subset of $\mathbb{R}^n$:
\begin{list}{}{
  \setlength\labelwidth{2.5em}
  \setlength\leftmargin{2.5em}
  }
  \item [(S1)]{\it i.i.d. entries}: Elements of ${\bf x}$ are i.i.d.;
  \item [(S2)]{\it finite variance}: The variance of $x_i$ is
    $\sigma_X^2 < \infty$ for all $i$.
\end{list}

These stochastic signal assumptions sometimes are referred to as
Bayesian signal model, and are commonly used in the literature \cite{Reeves2008,
Sarvotham2006, Reeves2013, Aeron2010}.
In addition, sparsity assumption, i.e., ${\bf x}$ is an
$s$-sparse signal, is sometimes adopted by using a specific
distribution \cite{Aeron2010, Reeves2013}.
In this paper, we consider the general signal that is sparse in some
orthonormal basis, instead of the signal that is sparse only in the identity
basis.
Thus, the sparsity assumption is not necessary. 

\subsection{Samples Assumptions}
\begin{figure}[t]
  \centering
  \includegraphics[width=0.7\textwidth]{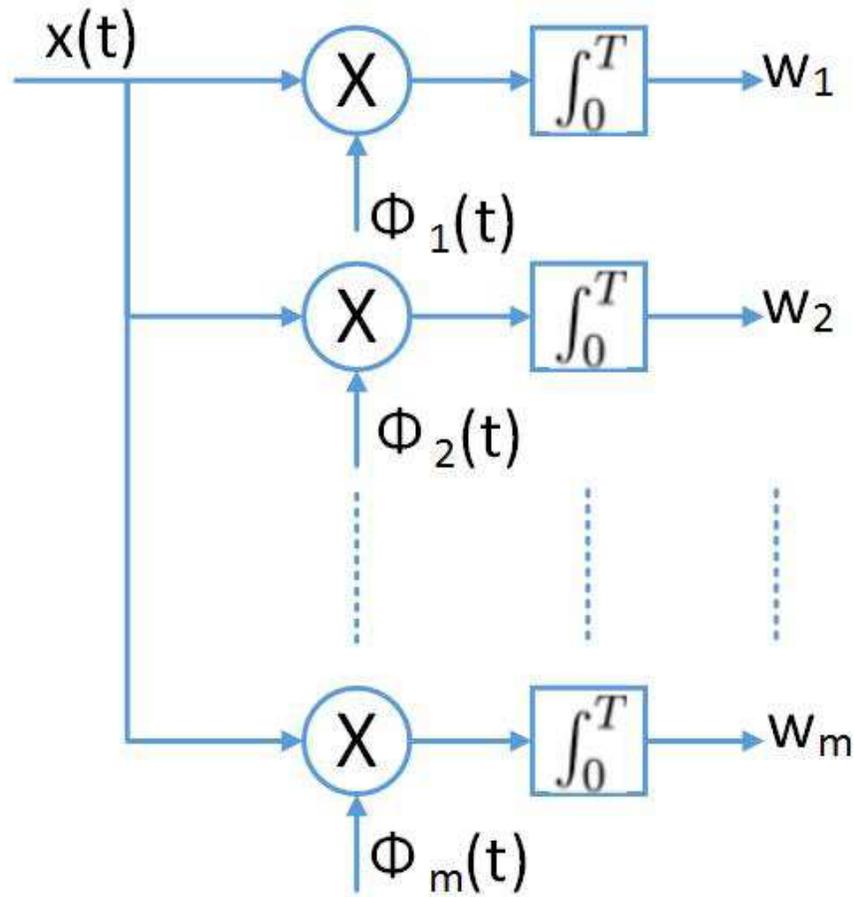}
  \caption{The structure of the AIC based on RMPI.}
  \label{fig:aic}
\end{figure}

A practical application of CS is the AIC which avoids high rate sampling
\cite{Candes2008, Laska2007}.
The structure of the AIC based on the random modulation pre-integration (RMPI)
is proposed in \cite{Candes2008}, as shown in \figurename\,\ref{fig:aic}.
Here the signal ${\bf x}(t)$ is an analog signal, and each waveform
${\bf \phi}_i(t)$ corresponds to a row in the sampling matrix
${\bf \Phi}$.
The AIC device consists of several parallel BMIs.
In each BMI, the analog signal is multiplied to a random sampling waveform ${\bf
\phi}_i(t)$ and then is integrated over the sampling period $T$.
Obviously, in the AIC shown in \figurename\,\ref{fig:aic}, the number of
samples is equal to the number of BMIs.

In the segmented CS architecture \cite{Taheri2011}, the sampling matrix ${\bf
\Phi}$ can be divided into two parts, i.e.,
\begin{align*}
  {\bf \Phi} = \left[ 
  \begin{array}{c}
    {\bf \Phi}_o \\
    {\bf \Phi}_e
  \end{array}
  \right]
\end{align*} 	
where ${\bf \Phi}_o \in \mathbb{R}^{m_o \times n}$ is the original
part, i.e., a set of original uncorrelated sampling
waveforms, and ${\bf \Phi}_e \in \mathbb{R}^{m_e \times n}$ is the
extended part.
Here $m = m_o+m_e$, with $m_o$ and $m_e$ being the number of
original samples and the number of additional samples, respectively.
Thus, the noiseless sample vector ${\bf w}$ can also be divided
into two parts, i.e., 
\begin{align*}
  {\bf w} = \left[
  \begin{array}{c}
    {\bf w}_o \\ 
    {\bf w}_e 
  \end{array}
  \right]
\end{align*}
where ${\bf w}_o = {\bf \Phi}_o {\bf x}$ and ${\bf w}_e =
{\bf \Phi}_e {\bf x}$ are the original sample and
additional sample vectors, respectively.
In ${\bf w}_o$, we have $m_o$ original samples, and in ${\bf w}_e$, we
have $m_e$ additional samples.
In practice, there are $m_o$ BMIs and the integration period $T$ is split into
$t$ sub-periods\cite{Taheri2011}.
Each BMI represents a row of ${\bf \Phi}_o$, and it outputs a sub-sample at
the end of every sub-period.
Hence, we can obtain $tm_o$ sub-samples during $t$ sub-periods from the
$m_o$ BMIs. 
With all these $tm_o$ sub-samples, we can construct $m_o$ original
samples in ${\bf w}_o$ and $m_e$ additional samples in ${\bf w}_e$ as
follows.

An original sample in ${\bf w}_o$ is generated by accumulating $t$
sub-samples from a single BMI.
Thus, the $m_o$ BMIs result in $m_o$ original samples in ${\bf w}_o$.
For each additional sample in ${\bf w}_e$, we consider a virtual BMI, which
represents a row of ${\bf \Phi}_e$.
At the end of every sub-period, the virtual BMI outputs one of the $m_o$
sub-samples from the $m_o$ real BMIs, and thus, after $t$ sub-periods, an
additional sample can be generated by accumulating $t$ sub-samples
over the $t$ sub-periods.
It is required that for each virtual BMI, the $t$ sub-samples are all
taken from different real BMIs (i.e., no two sub-samples are taken from
the same real BMI).
Thus, it is required that $t \leq m_o$.

\begin{figure}[t]
  \centering
  \includegraphics[width=0.7\textwidth]{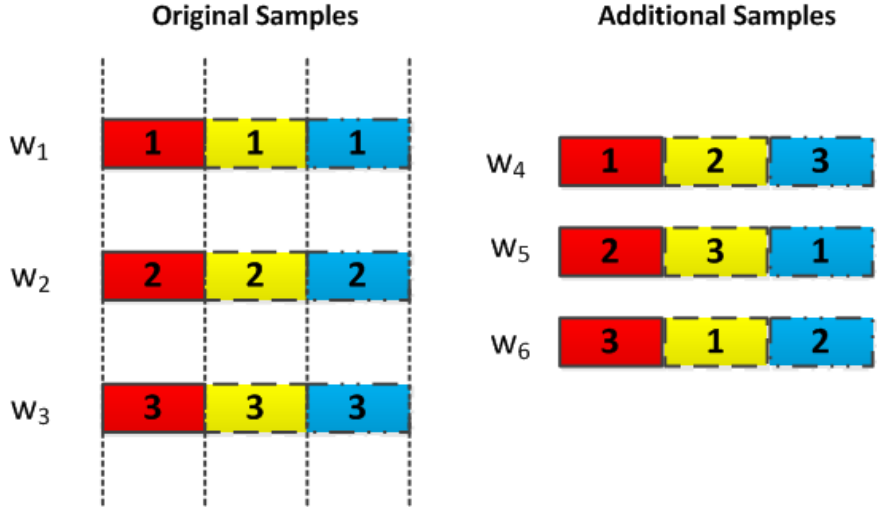}
  \caption{Construction of additional samples.}
  \label{fig:add_msr}
\end{figure}

\begin{myex}
  When $m_o=3$ and the integration period $T$ is divided into 3
  sub-periods, \figurename\,\ref{fig:add_msr} illustrates how
  additional samples are constructed.
  In \figurename\,\ref{fig:add_msr}, sub-samples are represented
  by rectangle boxes, and their corresponding sub-periods are represented by
  the colors of the rectangle boxes: red, yellow, and blue colors
  mean the first, second, and the third sub-periods, respectively.
  We have three original samples: $w_1, w_2$, and $w_3$.
  Each original sample consists of three sub-samples from the same
  real BMI.
  The number inside the ``sub-sample'' box indicates the index of
  the original sample (the index of the real BMI) that it comes from.
  We have the following observations on the additional samples $ w_4,
  w_5$ and $w_6$. 
  \begin{itemize}
    \item Each additional sample consists of 3
      sub-samples with different indices, which means that
      the sub-samples are selected from 3 different real BMIs.

    \item The order of the sub-samples in each additional
      sample is red, yellow and blue.
      It means that the $i$-th ($i=1, 2, 3$) sub-sample in an additional
      sample comes from the $i$-th sub-sample of an original
      sample, which is the output of the corresponding real BMI for
      the $i$-th sub-period.
  \end{itemize}
\end{myex}

From the above description, it can be seen that only $m_o$ parallel
BMIs are needed in the segmented CS-based AIC device, and $m$ ($\geq m_o$)
samples in total can be collected.
This implementation is equivalent to collecting additional samples
by multiplying the signal with additional sampling waveforms which are
not present among the actual BMI sampling waveforms, but rather each of theses
additional sampling waveforms comprises non-overlapping
sub-periods of different original waveforms.

Consider the following assumptions on the sampling matrix ${\bf
\Phi} \in \mathbb{R}^{m \times n}$.
The assumptions are
\begin{list}{}{
  \setlength\labelwidth{2.8em}
  \setlength\leftmargin{2.8em}
  }
  \item [(M1)]{\it non-adaptive samples}: The distribution of ${\bf
    \Phi}$ is independent of the signal ${\bf x}$ and the noise ${\bf
    z}$;
  \item [(M2)]{\it finite sampling rate}: The sampling rate
    $\delta$ is finite;
  \item [(M3)]{\it identically distributed}: Elements of ${\bf \Phi}$
    are identically distributed; 
  \item [(M4)]{\it zero mean}: The expectation of $\phi(i,j)$ is 0, where
    $\phi(i,j)$ denotes the $(i,j)$-th element of ${\bf \Phi}$;
  \item [(M5)]{\it finite variance}: The variance of $\phi(i,j)$ is $1/n$;
  \item [(M6)]{\it independent entries of ${\bf \Phi}_o$}: Elements of
    ${\bf \Phi}_o$ are independent;
  \item [(M7)]{\it uniform segment length}: Each row of ${\bf \Phi}$ can
    be divided into $m_o$ segments of length $l$, i.e., $l = n/m_o$ is
    an integer; the $i$-th ($i \in \{1, 2, \dots, m_o\}$) segment is
    corresponding to the $i$-th sub-period discussed before;
  \item [(M8)]{\it one-segment correlation}: For each row of ${\bf \Phi}_e$,
    the $i$-th ($i \in \{1, 2, \dots, m_o\}$) segment is copied from the
    $i$-th segment of a row of ${\bf \Phi}_o$, while ensuring that each row of
    ${\bf \Phi}_o$ contributes exactly one segment to each row of ${\bf
    \Phi}_e$.
    In other words, each row of ${\bf \Phi}_e$ is correlated to each
    row of ${\bf \Phi}_o$ over one segment only.\footnote{Here, for any two
    rows in ${\bf \Phi}$, if they have a common segment in a sub-period, we
    say the two rows are correlated over that segment/sub-period.}
\end{list}

In this paper, we consider general assumptions, i.e., the
assumptions (M1)--(M6) on the sampling matrix, which have also been
used, for example, in \cite{Reeves2013}.
Random Gaussian matrix is a specific example of the sampling matrix
satisfying the assumptions (M1)--(M6), and it has been used for the
information theoretic analysis on sparsity recovery or CS in some
other works \cite{Akcakaya2010, Wainwright2009, Wainwright2009a,
Aeron2010, Reeves2008}.
Actually, the assumptions (M1)--(M6) reflect the setting that the
samples are random projections of the signal, and the original
samples in ${\bf w}_o$ are uncorrelated.
In addition, the assumptions (M7) and (M8) characterize the
segmented CS architecture \cite{Taheri2011}.
Specifically, the integration period $T$ is equally divided
into several sub-periods, as suggested by the assumption (M7).
We further assume in the assumption (M7) that in each
sample, the number of
sub-periods/segments is also $m_o$, which is the same as the number of BMIs
and also the number of original uncorrelated samples.
As described before, the $i$-th sub-sample in an
additional sample comes from the $i$-th sub-sample of an
original sample.
This feature of the segmented CS-based AIC device
is reflected in the assumption (M8).

Based on these assumptions (especially assumptions (M7)
and (M8)), it can be seen that each row of ${\bf \Phi}_e$ (as well as
each additional sample in ${\bf w}_e$) actually corresponds to a
permuted sequence of $(1, 2, \dots, m_o)$, depending on the source BMI indices
of the $m_o$ segments of the row of ${\bf \Phi}_e$.
For example, as shown in \figurename\,\ref{fig:add_msr}, additional
sample ${\bf w}_5$ corresponds to the sequence $(2, 3, 1)$, which means
the first, the second and the third sub-samples of ${\bf w}_5$ come from the
second, the third and the first BMIs, respectively. 
Thus, there are at most $m_o!$ rows in ${\bf \Phi}_e$ and we have the
following observation on the $m_o!$ potential rows.
\begin{mylemma}
  \label{remark:add_row_group}
  These $m_o!$ potential rows can be divided into $(m_o-1)!$ groups,
  where each group consists of $m_o$ uncorrelated rows.
  \label{remark:group1}
\end{mylemma}
\begin{IEEEproof}
  Here we give an example of such grouping scheme.
   
  Since each of the $m_o!$ rows corresponds to a permuted sequence of
  $(1, 2, \dots, m_o)$, we need to prove that the
  $m_o!$ possible permuted sequences (including the sequence $(1, 2,
  \dots, m_o)$ itself) can be divided into $(m_o-1)!$ groups, and in
  each group, we have $m_o$ sequences in which any two sequences do not have
  correlation.\footnote{Recall that each potential row (for ${\bf \Phi}_e$)
  corresponds to a permuted sequence of $(1, 2, \dots, m_o)$.
  For any two rows, if they are correlated over the $j$-th
  segment/sub-period, the two sequences of the two rows have the same
  element at the $j$-th position.
  Accordingly, we say the two sequences are correlated over the $j$-th position.}

  First, among the $m_o!$ sequences, we consider those sequences
  whose first element is 1. 
  There are $(m_o-1)!$ such sequences.
  We put those $(m_o-1)!$ sequences in $(m_o-1)!$ groups, with each group
  having one sequence.
  Then, in each group, we perform cyclic shift on the corresponding sequence
  and we can generate $m_o-1$ new sequences by performing cyclic shift $m_o-1$
  times.
	In other words, in each time we move the final entry in the sequence to the
	first position, while shifting all other entries to their next positions.
  So in each group, we have $m_o$ sequences now, and the $m_o$ sequences are
  uncorrelated.
  It can be seen that: 1) totally there are $m_o!$ sequences in the $(m_o-1)!$
  groups; 
  2) in each group, any two sequences are different; 
  3) any two sequences from two different groups are different.
  Therefore, the above grouping satisfies Lemma~1.
  This completes the proof.
\end{IEEEproof}

\begin{mylemma}
  If $m_o$ is a prime, we can find $(m_o-1)$ groups from the $(m_o-1)!$ groups
  constructed as in Lemma~\ref{remark:group1} such that any two rows from
  different groups are correlated over one and only one segment.
  \label{remark:group2}
\end{mylemma}
\begin{IEEEproof}
  Throughout the proof, we establish the mapping from row to
  sequence as described in the proof of Lemma~\ref{remark:group1}.
  Consider $(m_o-1)$ sequences as follows: 
  in the $i$-th sequence $\mathcal{ R}_i$ ($i=1,2,...,m_o-1$), the $k$-th
  element ($k=1,2,...,m_o$) is $[1+(k-1)i]$ mod $m_o$.
  It is obvious that these $(m_o-1)$ sequences belong to $(m_o-1)$ different
  groups, and they are correlated over the first element only.
  Let the $i$-th sequence $\mathcal{R}_i$ belong to the $i$-th group denoted
  as $\mathcal{G}_i$.
  As shown in Lemma~\ref{remark:group1}, the rest ($m_o-1$) sequences in
  group $\mathcal{G}_i$ can be obtained by performing cyclic shift on
  $\mathcal{R}_i$.
  Therefore, in $\mathcal{G}_i$, for the sequence whose first element is $j$,
  the $k$-th element of the sequence can be expressed as $[j+(k-1)i]$ mod
  $m_o$ ($j, k=1, 2, \dots, m_o$).
  
  For any pair of $(i, j)$ and $(i', j')$ where $i \neq i'$, $i, i' \in \{1, 2, \dots,
  m_o-1\}$ and $j, j' \in \{1, 2, \dots, m_o\}$, the greatest common divisor
  of ($i-i'$) and $m_o$, denoted GCD($i-i', m_o$), is 1 since $m_o$ is a prime.
  Therefore, we have $-(j-j') = 0$ mod GCD($i-i', m_o$).
  Then according to the linear congruence equation, with given pair of $(i, j)$
  and $(i', j')$, the equation 
  \begin{equation*}
    (k-1)(i-i') = -(j-j') \quad {\rm mod} \quad m_o
  \end{equation*}
  has an unique solution 
  $k^* \in \{1, 2, \dots, m_o\}$ \cite{LinearCongEqMathworld}.
  In other words, we can always find one and only one $k^* \in \{1, 2, \dots,
  m_o\}$ that makes $[j+(k^*-1)i] = [j'+(k^*-1)i']$ mod $m_o$.
  Therefore, for any two sequences from two different groups $\mathcal{G}_i$
  and $\mathcal{G}_{i'}$, they are correlated over exactly one position.
  This completes the proof.
\end{IEEEproof}

Define the {\it extension rate of the CS system} with correlated
samples as the ratio of the number of additional samples
$m_e$ to the number of original samples $m_o$, i.e., 
$\alpha \overset{\triangle}{=} m_e / m_o$.
In this paper, we consider two kinds of ${\bf \Phi}_e$:
\begin{list}{}{
  \setlength\labelwidth{3.5em}
  \setlength\leftmargin{3.5em}
  }
  \item [(M9a)] ${\bf \Phi}_e$ consists of $m_e$ rows with $m_e
    \leq m_o$;
    all these $m_e$ rows are uncorrelated, and are taken from one of the
    $(m_o-1)!$ groups of potential rows constructed as shown in Lemma~\ref{remark:group1}; 
    in this case, $\alpha \leq 1$; 
    
  \item [(M9b)] ${\bf \Phi}_e$ consists of all rows in $\alpha$ groups of potential
    rows constructed as shown in Lemma~\ref{remark:group2}; in this case,
    $\alpha = 1, 2, \dots, m_o-1$.
\end{list}

\section{Main results}
\label{sec:main_results}
The channel capacity $C$ of the CS system (see
\figurename\,\ref{fig:block_diag}) is studied in this section.
The channel capacity in the considered setup gives the amount of information that
can be extracted from the compressed samples.
Meanwhile, the rate-distortion function $R(D)$ gives the
minimum information (in bits) needed to reconstruct the signal with
distortion $D$ for a given distortion measure.
Accordingly, an inequality between $C$ and $R(D)$ can be given using
the source-channel separation theorem \cite{Gamal2012}, which results in a lower bound
on the sampling rate $\delta$ as a function of distortion $D$ and SNR $\gamma$.
Apparently, when the CS system has correlated
samples, the amount of information that can be extracted from the
samples decreases.
In other words, the channel capacity $C$ is smaller than that of
the CS system in which all samples are uncorrelated.
Thus, we expect a penalty term in the upper bound on the
channel capacity $C$ and in the lower bound on the sampling rate
$\delta$. 
According to assumption (M8), in the sampling matrix ${\bf
\Phi}$, an additional row is correlated with an original row over one segment.
Thus, when the variance of the signal, the variance of the entries of ${\bf
\Phi}$, and the length of the segment are fixed, as assumed in (S2), (M5) and
(M7), respectively, the correlation between an additional sample and
an original sample is fixed.
The penalty term caused by the fixed correlation among samples is
discussed in the remaining part of this section.

\subsection{Case 1: $\alpha \leq 1$}
\label{ssec:result_alpha<1}
The following lemma gives a bound on the capacity of the
CS system with correlated samples and a sampling
matrix satisfying the assumption (M9a).
\begin{mythm}
  For a signal satisfying the assumptions (S1)--(S2) and a
  sampling matrix satisfying the assumptions (M1)--(M8) and
  (M9a), the maximal amount of information that can be extracted from
  the samples is given by
  \begin{equation}
    C \leq \frac{m}{2} \log(\gamma+1)
    + \frac{1}{2} \log \left[ 1-\left(\frac{\gamma}{\gamma+1}\right)^2
    \cdot \alpha \right]
    \label{ineq:capacity_corr_1}
  \end{equation}
  with equality achieved if and only if ${\bf w} \sim \mathcal{N}(0,
  {\bf \Sigma}_W)$.
  \label{lemma:capacity_corr_1}
\end{mythm}
\begin{IEEEproof}
  See
  Appendix~\ref{app:comm_proof}~and~then follow with
  Appendix~\ref{app:proof_alpha<1} for the proof.
\end{IEEEproof}

It can be observed that the second term on the right-hand-side of 
(\ref{ineq:capacity_corr_1}) is a
function of $\gamma$ and $\alpha$ and it is always non-positive.
Thus, this term has a meaning of the penalty term caused by the fixed
correlation among samples. 
Furthermore, if the total number of samples $m$ is fixed, the
upper bound in (\ref{ineq:capacity_corr_1}) is obviously decreasing
as $\alpha$ increases, which means that when the total number of
samples $m$ is fixed, it is better to have less
correlated samples.  
However, usually the number of original samples $m_o$
(not the total number of samples $m$) is fixed, and we are
interested in the best extension rate $\alpha$. 
Since $m = (1+\alpha)m_o$, (\ref{ineq:capacity_corr_1}) becomes
\begin{equation}
  C \leq \frac{(1\!+\!\alpha)m_o}{2} \log(\gamma+1)
  + \frac{1}{2} \log \left[ 1\!-\!\left(\!\frac{\gamma}{\gamma+1}\!\right)^2
  \!\cdot\!\alpha \right].
  \label{ineq:capacity_corr_1a}
\end{equation}
The right-hand-side of (\ref{ineq:capacity_corr_1a}) is not always an
increasing function of $\alpha$.
However, noting that
$\alpha\overset{\triangle}{=}m_e/m_o$ where $m_e$ is an integer, we
have the following observation on the upper bound in
(\ref{ineq:capacity_corr_1a}).
\begin{mylemma}
  The maximum of the upper bound on $C$ in
  (\ref{ineq:capacity_corr_1a}) is achieved when $\alpha=1$ for all
  $m_o \geq 1$ and positive $\gamma$.
  \label{remark:opt_alpha}
\end{mylemma}
\begin{IEEEproof}
  Denote the right-hand-side of (\ref{ineq:capacity_corr_1a}) as
  $f(\alpha)$.
  Thus, the first-order derivative of $f(\alpha)$ is given by
  \begin{equation}
    f'(\alpha) = \frac{m_o}{2}\log(\gamma+1)
    - \frac{1}{2\ln 2} \frac{\gamma^2}{(\gamma+1)^2 - \gamma^2\alpha}.
    \label{}
  \end{equation}
  Obviously, $f'(\alpha)$ is a strictly decreasing function of
  $\alpha$ for $0\leq \alpha \leq 1$.
  Thus, $f'(1) \leq f'(\alpha) \leq f'(0)$, where
  \begin{align}
    f'(0) &= \frac{m_o}{2} \log(\gamma+1) - \frac{1}{2\ln 2}
    \frac{\gamma^2}{(\gamma+1)^2} \label{eq:gradf0}\\
    f'(1) &= \frac{m_o}{2} \log(\gamma+1) -
    \frac{1}{2\ln 2}\frac{\gamma^2}{2\gamma+1}.
  \end{align}

  We first show that $f'(0) > 0$.
  Let $g(\gamma) = \ln(\gamma+1) - \gamma / (\gamma+1)$.
  The derivative of $g(\gamma)$ is
  \begin{align*}
    g'(\gamma) = \frac{1}{\gamma+1} - \frac{1}{(\gamma+1)^2}
    = \frac{\gamma}{(\gamma+1)^2}
    > 0.
  \end{align*}
  Thus, $g(\gamma) > g(0) = 0$ and we have 
  $\ln(\gamma+1) > \gamma / (\gamma+1)$ for $\gamma > 0$.
  Using (\ref{eq:gradf0}), we then have the following inequality
  \begin{align}
    f'(0)
    &> \frac{m_o}{2\ln 2} \frac{\gamma}{\gamma+1}
    - \frac{1}{2\ln 2} \frac{\gamma^2}{(\gamma+1)^2} \nonumber \\
    &\geq \frac{1}{2\ln 2} \frac{\gamma}{\gamma+1}
    - \frac{1}{2\ln 2} \frac{\gamma^2}{(\gamma+1)^2} \label{ineq:f'(0)_1}\\
    &= \frac{1}{2\ln2} \frac{\gamma}{(\gamma+1)^2} > 0 \nonumber
  \end{align}
  where the second inequality follows from $m_o \geq 1$.

  If $\alpha$ can be chosen from a continuous set between
  0 and 1, $f'(\alpha)$ is a strictly decreasing function of $\alpha$.
  Note that $f'(0) > 0$. 
  Therefore, when $f'(1) \geq 0$, the maximum of $f(\alpha)$ is achieved at
  $\alpha_1=1$; when $f'(1) < 0$, the maximum of $f(\alpha)$ is achieved at 
  $\alpha_2 = (\gamma+1)^2/\gamma^2 - 1/[m_o \ln(\gamma+1)]$,
  which makes $f'(\alpha)=0$.

  Since $m_o \geq 1$, we have 
  \begin{align}
    &\quad\ m_o\alpha_2 - (m_o-1) \nonumber \\
    &= m_o \left(\frac{\gamma+1}{\gamma}\right)^2 -
    \frac{1}{\ln(\gamma+1)} - (m_o-1) \label{ineq:opt_alpha_0} \\
    &\geq \left(\frac{\gamma+1}{\gamma}\right)^2 -
    \frac{1}{\ln(\gamma+1)} \label{ineq:opt_alpha_1}\\
    &> \left(\frac{\gamma+1}{\gamma}\right)^2 -
    \frac{\gamma+1}{\gamma} \label{ineq:opt_alpha_2}\\
    &= \frac{\gamma+1}{\gamma^2} > 0\nonumber
  \end{align}
  where (\ref{ineq:opt_alpha_1}) follows from the fact that
  (\ref{ineq:opt_alpha_0}) is an increasing function of $m_o$,
  and (\ref{ineq:opt_alpha_2}) follows from the inequality $\ln(\gamma+1) >
  \gamma/(\gamma+1)$.
  Therefore, $\alpha_2 \geq (m_o-1)/m_o$.
  Noting that $\alpha$ can only take a value from the discrete set
  $\{0, 1/m_o, 2/m_o, \dots, 1\}$, when $f'(1) \geq 0$, the
  maximum of $f(\alpha)$ is achieved at $\alpha_1=1$; when $f'(1) < 0$, the
  maximum of $f(\alpha)$ is either $f(\alpha_1)$ or $f(\alpha_3)$, whichever
  is larger.
  Here $\alpha_3=(m_o-1)/m_o$. 
  We have
  \begin{align}
    &\quad\ f(\alpha_1) - f(\alpha_3) \nonumber \\
    &= m_o \log(\gamma+1) + \frac{1}{2} \log
    \left[1-\left(\frac{\gamma}{\gamma+1}\right)^2 \right] \nonumber \\
    &\quad - \frac{2m_o-1}{2} \log(\gamma+1) - \frac{1}{2} \log
    \left[1-\left(\frac{\gamma}{\gamma+1}\right)^2 \cdot
    \frac{m_o-1}{m_o} \right] \nonumber \\
    &= \frac{1}{2} \log(\gamma+1) 
    + \frac{1}{2}\log\left[1-\left(\frac{\gamma}{\gamma+1}\right)^2\right]
    \nonumber \\
    &\quad - \frac{1}{2}\log\left[1-\left(\frac{\gamma}{\gamma+1}\right)^2
    \cdot \frac{m_o-1}{m_o} \right]. \label{ineq:opt_alpha_4}
  \end{align}
  Since $\log\left[1-\left(\gamma / (\gamma+1)\right)^2
  \cdot (m_o-1)/m_o \right] \leq 0$ considering $m_o \geq 1$, we have
  \begin{align*}
    &\quad\ f(\alpha_1) - f(\alpha_3) \nonumber \\
    &\geq \frac{1}{2} \log(\gamma+1) 
    + \frac{1}{2}\log\left[1-\left(\frac{\gamma}{\gamma+1}\right)^2\right]
    \label{ineq:opt_alpha_5} \\
    &= \frac{1}{2} \log \left[ \frac{2\gamma+1}{\gamma+1} \right]
    > 0.
  \end{align*}
  Thus, $f(1) > f((m_o-1)/m_o)$.
  In other words, the maximum of $f(\alpha)$ is always achieved when
  $\alpha=1$.
  This completes the proof.
\end{IEEEproof}

Based on Theorem~\ref{lemma:capacity_corr_1}, a lower bound on the
sampling rate $\delta$ is given in the following theorem. 

\begin{mythm}
  For a signal satisfying the assumptions (S1)--(S2) and a
  sampling matrix satisfying the assumptions (M1)--(M8) and (M9a), if a distortion
  $D$ is achievable, then
  \begin{equation}
    \delta \geq \frac{2R(D)}{\log(\gamma+1)}
    -\frac{1}{n}\frac{\log \left[1 -
    \left(\frac{\gamma}{\gamma+1}\right)^2
    \cdot \alpha \right]}{\log(\gamma+1)} 
    \label{ineq:thm1}
  \end{equation}
  as $n \rightarrow \infty$.
  \label{thm:msr_rate_1}
\end{mythm}
\begin{IEEEproof}
  According to the source-channel separation theorem for discrete-time
  continuous amplitude stationary ergodic signals, ${\bf x}$ can be
  communicated up to distortion $D$ via several channels if and only
  if the information content $C$ that can be extracted from these
  channels exceeds the information content $nR(D)$ of the signal ${\bf
  x}$ \cite{Berger1971}.
  In other words, $nR(D) \leq C$ when $n$ goes to $\infty$.
  According to Theorem~\ref{lemma:capacity_corr_1}, the information
  content $C$ is upper bounded by (\ref{ineq:capacity_corr_1}).
  Meanwhile, $nR(D)$ gives the minimal number of bits in the $n$ source
  symbols in ${\bf x}$ needed to recover ${\bf x}$ within
  distortion $D$.
 
  Therefore, we have
  \begin{align*}
    nR(D) \leq C \leq \frac{m}{2} \log(\gamma+1)
    + \frac{1}{2} \log \left[ 1-\left(\frac{\gamma}{\gamma+1}\right)^2
    \cdot \alpha \right]
  \end{align*}
  which implies that
  \begin{align*}
    \delta = \frac{m}{n} \geq 
    \frac{2R(D)}{\log(\gamma+1)}
    -\frac{1}{n}\frac{\log \left[1 -
    \left(\frac{\gamma}{\gamma+1}\right)^2
    \cdot \alpha \right]}{\log(\gamma+1)}.
  \end{align*}
 
  This completes the proof.
\end{IEEEproof}

If $\alpha=0$, i.e., all samples are uncorrelated, the result
is essentially the same as that in \cite{Sarvotham2006}.
If $\alpha > 0$, the second term on the right-hand-side of
(\ref{ineq:thm1}), which is the penalty term, is vanishing as
$n~\rightarrow~\infty$.
In other words, the penalty because of the fixed
correlation among samples vanishes as $n~\rightarrow~\infty$.

Note that the original sampling rate 
$\delta_o \overset{\triangle}= m_o/n = \delta/(1+\alpha)$
is the parameter to be designed for a segmented CS-based AIC device. 
Thus, it is interesting to study how the extension rate $\alpha$ affects the
required $\delta_o$ in order to achieve a given distortion level.
In terms of $\delta_o$, the inequality (\ref{ineq:thm1}) becomes
\begin{equation}
  \delta_o \geq \frac{1}{1+\alpha}\left(\frac{2R(D)}{\log(\gamma+1)}
  \!-\!\frac{1}{n}\frac{\log
  \left[1\!-\!\left(\frac{\gamma}{\gamma+1}\right)^2
  \!\cdot\!\alpha \right]}{\log(\gamma+1)}\right).
  \label{ineq:org_msr_rate}
\end{equation}

Although the optimal $\alpha$ that minimizes the
right-hand-side of (\ref{ineq:org_msr_rate}) is not easy to
find out from this expression, it
can still be observed that as $n \rightarrow \infty$, the
right-hand-side of (\ref{ineq:org_msr_rate}) becomes a strictly
decreasing function of $\alpha$, which means that the required
original sampling rate decreases as the extension rate $\alpha$
increases.
Considering that (\ref{ineq:org_msr_rate}) essentially corresponds to
(\ref{ineq:capacity_corr_1a}), Numerical Example~1 in
Section~\ref{sec:num_results} shows that the lower bound on $\delta_o$
behaves similar to the upper bound on $C$ in
(\ref{ineq:capacity_corr_1a}), and the minimum is achieved when
$\alpha=1$ since $\alpha$ can only take values from $0, 1/m_o, 2/m_o,
\dots, 1$.

\subsection{Case 2: $\alpha = 1, 2, \dots, m_o-1$}
\label{ssec:result_alpha>1}
The following lemma now gives a bound on the capacity of
the CS system with correlated samples and
a sampling matrix satisfying the
assumption (M9b).
This lemma extends the result of Theorem~\ref{lemma:capacity_corr_1}.
\begin{mythm}
  For a signal satisfying the assumptions (S1)--(S2) and a
  sampling matrix satisfying the assumptions (M1)--(M8) and
  (M9b), the maximal amount of information that can be extracted from
  the samples is given by
  \begin{equation}
    C \leq \frac{m}{2} \log (\gamma\!+\!1)
    - \left[\frac{\alpha\!+\!1}{2} \log (\gamma\!+\!1)
    - \frac{1}{2} \log \left( (1\!+\!\alpha)\gamma\!+\!1 \right) \right]
    \label{ineq:capacity_corr_2}
  \end{equation}
  with equality achieved if and only if ${\bf w} \sim \mathcal{N}(0,
  {\bf \Sigma}_W)$.
  \label{lemma:capacity_corr_2}
\end{mythm}
\begin{IEEEproof}
  See Appendix~\ref{app:comm_proof}~and then follow with
  Appendix~\ref{app:proof_alpha>1} for the proof.
\end{IEEEproof}

It can be observed that the terms in the square brackets on
the right-hand-side of (\ref{ineq:capacity_corr_2}) are the penalty
terms caused by the fixed correlation among samples.
We have the following observation on the upper bound on
$C$ in (\ref{ineq:capacity_corr_2}).
\begin{mylemma}
  Considering that $\alpha$ is an integer in $\{1, 2, \dots,
  m_o-1\}$, the
  upper bound on $C$ in (\ref{ineq:capacity_corr_2}) increases as
  $\alpha$ increases.
  \label{remark:C_ub_incr_alpha>1}
\end{mylemma}
\begin{IEEEproof}
  Denote the right-hand-side of (\ref{ineq:capacity_corr_2}) as
  $h(\alpha)$.
  Then we have
    \begin{equation*}
    h(\alpha+1) - h(\alpha)
    = \frac{m_o-1}{2} \log(\gamma+1) 
    + \frac{1}{2} \log \left[
    \frac{1+(\alpha+2)\gamma}{1+(\alpha+1)\gamma}
    \right].
  \end{equation*}
  Since $(1+(\alpha+2)\gamma) / (1+(\alpha+1)) \geq 1$, $m_o \geq 1$, and 
  $\gamma \geq 0$, we have $h(\alpha+1) \geq h(\alpha)$.
  This completes the proof.
\end{IEEEproof}

When $\alpha=1$, the result in (\ref{ineq:capacity_corr_2}) is the
same as that in (\ref{ineq:capacity_corr_1}).
From the proof of Lemma~\ref{remark:opt_alpha} we can see
that the upper bound on $C$ in
(\ref{ineq:capacity_corr_1}) is an increasing function of $\alpha$
when $\alpha$ takes values from $\{0, 1/m_o, 2/m_o, \dots, 1\}$,
and thus the maximum of the upper bound on $C$ in
(\ref{ineq:capacity_corr_1}) is achieved when $\alpha = 1$.
From Lemma~\ref{remark:C_ub_incr_alpha>1}, the minimum of the upper
bound on $C$ in (\ref{ineq:capacity_corr_2}) is achieved when
$\alpha=1$.
Thus, the upper bound on $C$ in (\ref{ineq:capacity_corr_2}) is always
higher than that in (\ref{ineq:capacity_corr_1}).
This is reasonable because when the number of original uncorrelated
samples $m_o$ is fixed, more correlated samples can be taken
with assumption (M9b) than that with assumption (M9a).

Based on Theorem~\ref{lemma:capacity_corr_2}, a lower bound on the
sampling rate $\delta$ is given in the following theorem.

\begin{mythm}
  For a signal satisfying the assumptions (S1)--(S2) and a
  sampling matrix satisfying the assumptions (M1)--(M8) and
  (M9b), if a distortion $D$ is achievable, then
  \begin{equation}
    \delta \geq \frac{2R(D)}{\log(\gamma+1)}
    +\frac{\alpha+1}{n} 
    -\frac{1}{n} \frac{\log [(1+\alpha)\gamma+1]}{\log(\gamma+1)}
    \label{ineq:thm2}
  \end{equation}
  as $n \rightarrow \infty$.
  \label{thm:msr_rate_2}
\end{mythm}
\begin{IEEEproof}
  The proof follows the same steps as that of
  Theorem~\ref{thm:msr_rate_1}.
\end{IEEEproof}

In this case, the penalty brought by the fixed correlation between
samples also vanishes as $n \rightarrow \infty$.

Similar to the case of $\alpha \leq 1$, the original
sampling rate $\delta_o$ satisfies
\begin{align}
  \delta_o \geq \frac{1}{1+\alpha}
  \left(
  \frac{2R(D)}{\log(\gamma+1)}
  -\frac{1}{n} \frac{\log [(1+\alpha)\gamma+1]}{\log(\gamma+1)}
  \right)
  + \frac{1}{n}.
  \label{ineq:org_msr_rate_2}
\end{align}
As $n \rightarrow \infty$, the right-hand-side of
(\ref{ineq:org_msr_rate_2}) becomes a strictly decreasing function of
$\alpha$, which means that the required original sampling rate
decreases as the extension rate $\alpha$ increases.

\begin{figure}[t]
  \centering
  \includegraphics[width=0.8\textwidth]{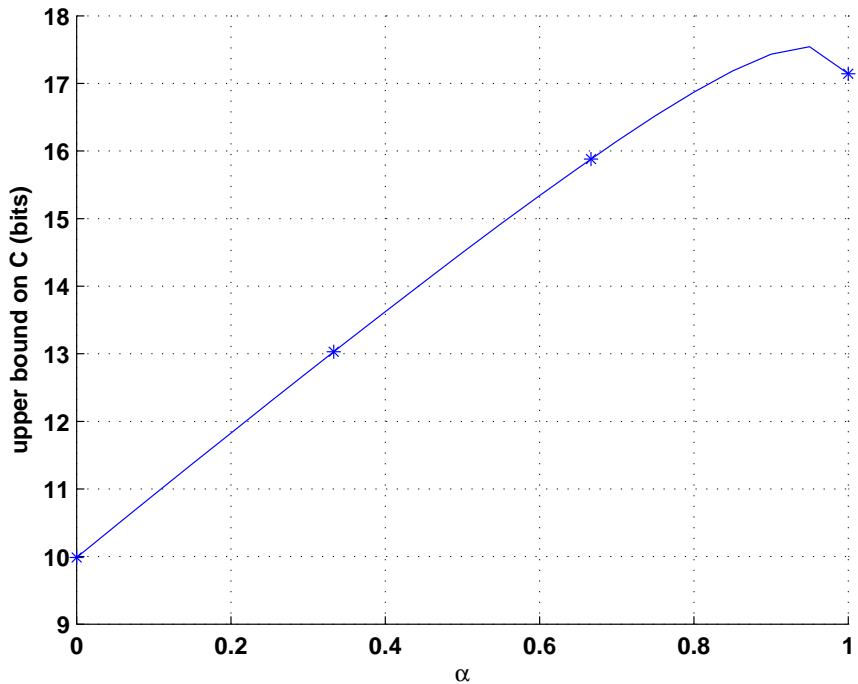}
  \caption{Upper bound on the capacity $C$ in
  (\ref{ineq:capacity_corr_1a}) versus $\alpha$.}
  \label{fig:capacity_ub}
\end{figure}

\begin{figure}[t]
  \centering
  \includegraphics[width=0.8\textwidth]{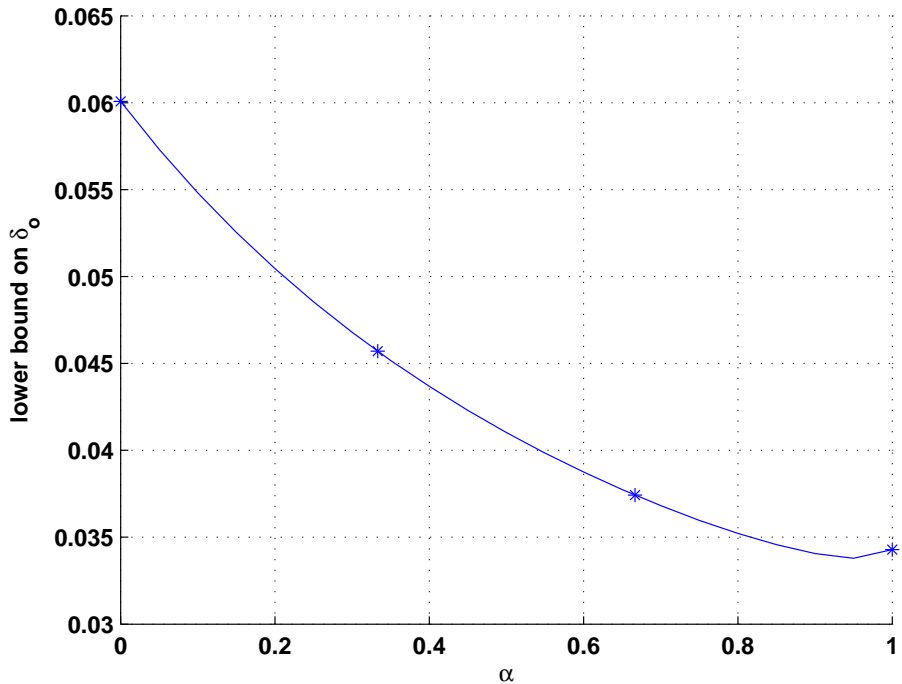}
  \caption{Lower bound on the original sampling rate $\delta_o$ in
  (\ref{ineq:org_msr_rate}) versus $\alpha$.}
  \label{fig:delta_o_lb}
\end{figure}

\section{Numerical Results}
\label{sec:num_results}
To illustrate Lemma~\ref{remark:opt_alpha}, we consider
the following example.
\begin{mysex}
  Consider the sampling matrix satisfying assumptions (M1)--(M8) and (M9a).
  Let the SNR $\gamma$ be 20~dB, the number of original samples
  $m_o$ be 3, and the signal's length $n$ be 100.
  The rate-distortion function $R(D)$ is 0.2 bits/symbol
  in the example.
\end{mysex}
Figs.\,\ref{fig:capacity_ub}~and~\ref{fig:delta_o_lb} show the
upper bound on $C$ in (\ref{ineq:capacity_corr_1a}) and the lower
bound on $\delta_o$ in (\ref{ineq:org_msr_rate}), respectively, for
different values of $\alpha$.
It can be observed from both figures that the optimum, i.e., the
maximum of the upper bound on $C$ (or the minimum of the lower bound
on $\delta_o$) is achieved at 
$\alpha = (\gamma+1)^2/\gamma^2 - 1/[m_o\ln(\gamma+1)] \approx 0.95$
if $\alpha$ can take any continuous value between 0 and 1.
However, considering that $\alpha$ can only take values
$0, 1/3, 2/3$ and $1$ in this example, as shown by the points marked
by `*' in both figures, the optimum is achieved at $\alpha=1$.
This verifies the Lemma~\ref{remark:opt_alpha}.

Next, we illustrate
Theorems~\ref{thm:msr_rate_1}~and~\ref{thm:msr_rate_2}.
\begin{mysex}
  Consider an $s$-sparse signal ${\bf x}$ where the spikes have
  uniform amplitude and the sparsity ratio $s/n$ is fixed as
  $10^{-4}$.
  In this case, it is well known that precise description of ${\bf x}$
  would require approximately $\log \binom{n}{s} \approx s\log(n/s)$
  bits \cite{Sarvotham2006}.
  Accordingly, $R(D)$ is approximately calculated as $(s/n)\log
  (n/s) = 0.0013$ bits/symbol.
  \label{ex2}
\end{mysex}

\begin{figure}[t]
  \centering
  \includegraphics[width=0.8\textwidth]{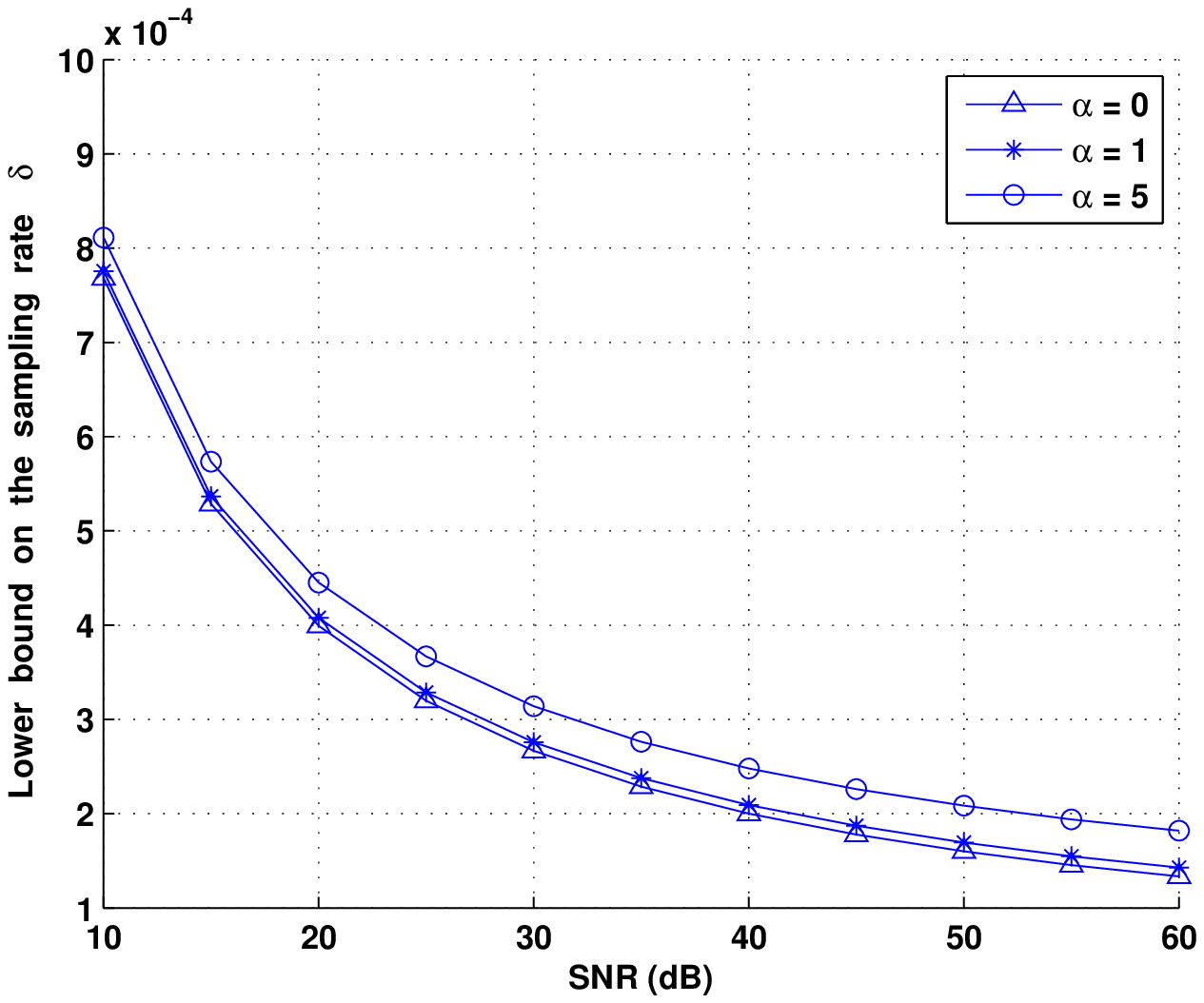}
  \caption{Lower bound on the sampling rate $\delta$ versus SNR for 
	different $\alpha=0,1,5$ when $n=10^5$.}
  \label{fig:delta_lb_1}
\end{figure}

\begin{figure}[t]
  \centering
  \includegraphics[width=0.8\textwidth]{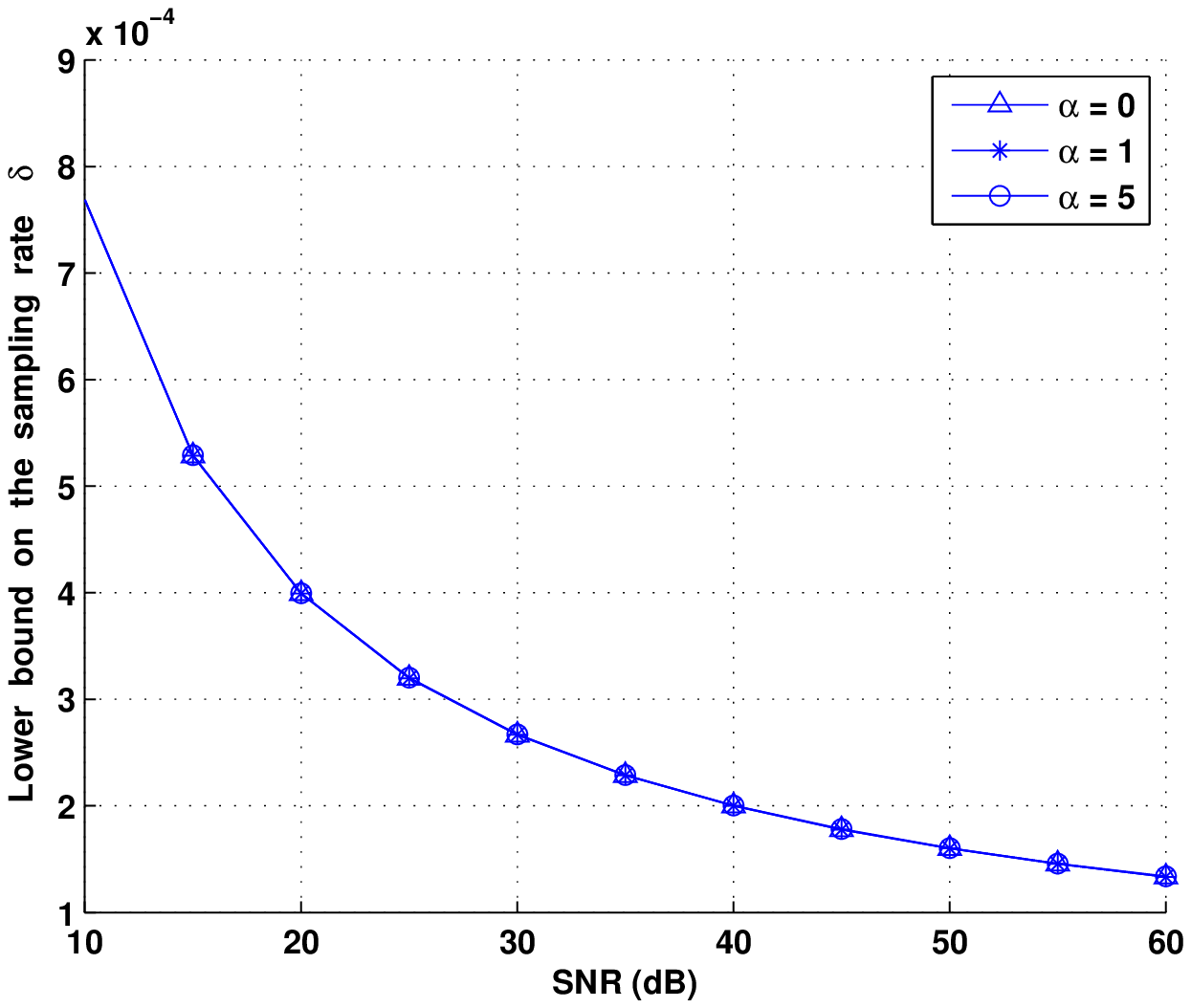}
  \caption{Lower bound on the sampling rate $\delta$ versus SNR for 
	different $\alpha=0,1,5$ when $n=10^7$.}
  \label{fig:delta_lb_2}
\end{figure}

In Figs.\,\ref{fig:delta_lb_1}~and~\ref{fig:delta_lb_2}, the lower
bounds on the sampling rate $\delta$ in
either (\ref{ineq:thm1}) or (\ref{ineq:thm2}) (based on the value of
$\alpha$) for $n=10^5$ and $n=10^7$ are shown.
It can be observed from both figures that as the SNR $\gamma$
increases, the lower bound on the sampling rate $\delta$ decreases,
which means that fewer samples are needed for a higher SNR.
Besides, as $\alpha$ increases, the lower bound on $\delta$ increases as well.
The gap between the curve with $\alpha=0$ and that with the other values
of $\alpha$ is the penalty brought by the fixed correlation among samples.
However, comparing
Figs.\,\ref{fig:delta_lb_1}~and~\ref{fig:delta_lb_2} to each other, it
can be seen that this penalty vanishes as $n$ increases, which verifies
the conclusions obtained based on
Theorems~\ref{thm:msr_rate_1}~and~\ref{thm:msr_rate_2}.

\begin{mysex}
  Continuing with the same setup as used in Example~\ref{ex2}, let
  $n=10^7$.
  In the segmented CS architecture, the additional samples ${\bf
  \Phi}_e {\bf x}$ can be obtained from the original samples ${\bf
  \Phi}_o {\bf x}$ \cite{Taheri2011}.
  Thus, in this example we show how the extension rate
  $\alpha$ affects the requirement on the original sampling rate
  $\delta_o$.
\end{mysex}

\begin{figure}[t]
  \centering
  \includegraphics[width=0.8\textwidth]{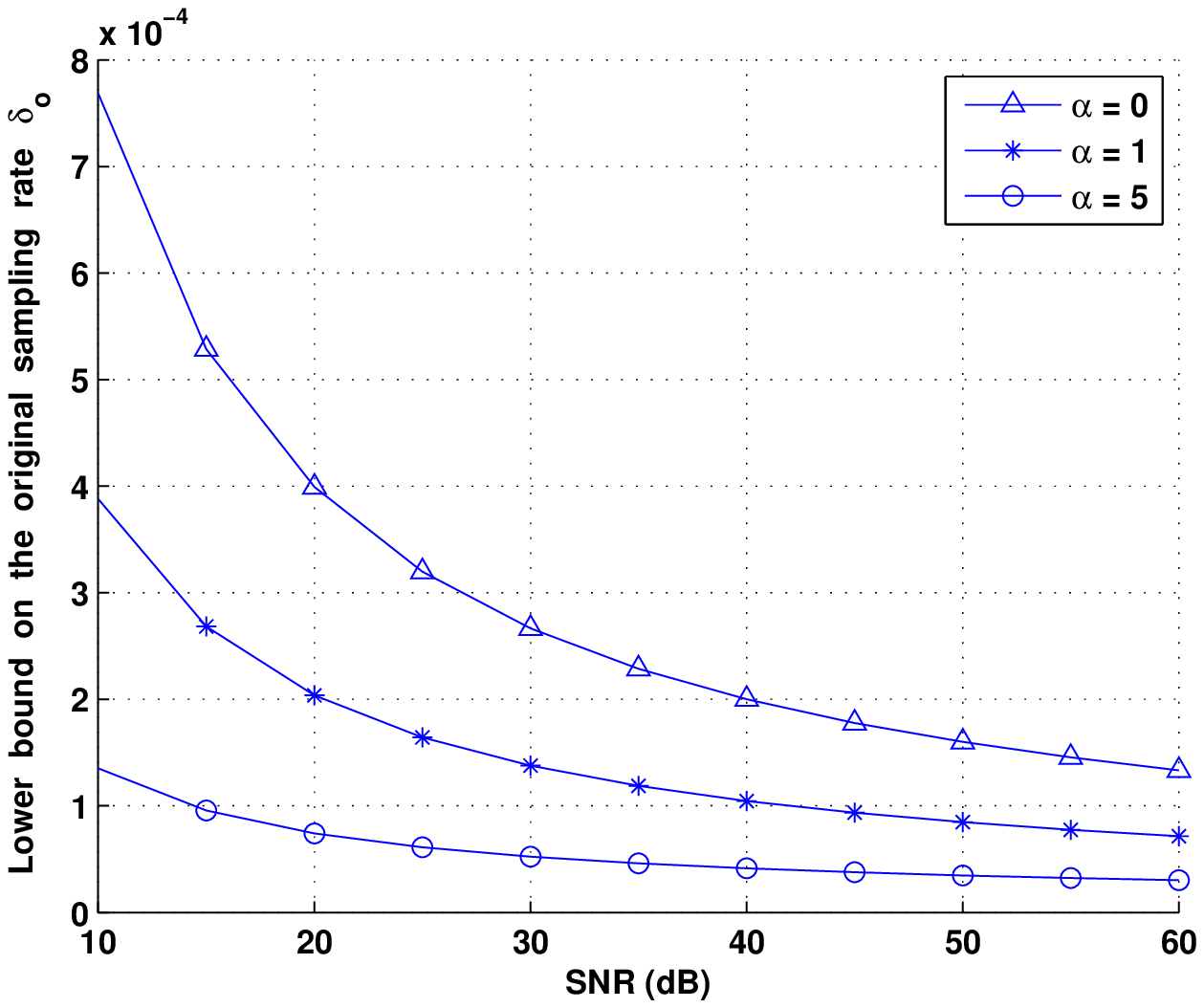}
  \caption{Lower bound on the original sampling rate $\delta_o$ versus 
	SNR for different $\alpha=0,1,5$ when $n=10^7$.}
  \label{fig:delta_o_lb_1}
\end{figure}

\figurename\,\ref{fig:delta_o_lb_1} shows the lower bound on the
$\delta_o$ in either (\ref{ineq:org_msr_rate}) or
(\ref{ineq:org_msr_rate_2}) (based on the value of $\alpha$) for
different extension rates $\alpha$.
It can be observed that as $\alpha$ increases, the lower bound on the
original sampling rate $\delta_o$ decreases, which means that fewer
original samples are needed to achieve the same reconstruction
performance.
This confirms and explains the advantage of using
segmented CS architecture over the non-segmented CS architecture one
of \cite{Laska2007}.

\section{Conclusion}
\label{sec:concl}
The performance limits of the segmented CS  have been studied where 
samples are correlated. When the total number of samples is fixed, 
there is a performance degradation brought by the fixed correlation 
among samples by segmented CS. This performance degradation is 
characterized by a penalty term in the upper bound on the channel 
capacity of the corresponding sampling matrix or in the lower bound 
on the sampling rate. This degradation is vanishing as the dimension 
of the signal increases, which has also been verified by the numerical 
results. From another point of view, as the extension rate increases, 
the necessary condition on the original sampling rate to achieve a 
given distortion level becomes weaker, i.e., fewer original samples 
(BMIs in the AIC) are needed. This verifies the advantages of the 
segmented CS architecture over the non-segmented CS one.

\appendices
\section{Common Start of Proof for Theorems~\ref{lemma:capacity_corr_1}
and~\ref{lemma:capacity_corr_2}}
\label{app:comm_proof}
The channel in \figurename\,\ref{fig:block_diag} can be
formalized as 
\begin{equation}
  {\bf y} = {\bf w} + {\bf z}.
  \label{eq:channel}
\end{equation}
The channel capacity is given as \cite{Cover1991}
\begin{equation}
  C = \max_{p_{WY} ({\bf w}, {\bf y})} I({\bf w}; {\bf y})
  \label{eq:capacity_mutual_info}
\end{equation}
where $p_{WY} ({\bf w}, {\bf y})$ denotes the joint probability of
two $m$-dimensional random vectors ${\bf w}$ and ${\bf y}$ and
$I({\bf w}; {\bf y})$ denotes the mutual information between two
random vectors ${\bf w}$ and ${\bf y}$.
Let $h(\cdot)$ denote the entropy of a random vector.
Then, the mutual information can be expressed as
\begin{align}
  I({\bf w}; {\bf y}) &= h({\bf y}) - h({\bf y}|{\bf w}) 
  = h({\bf y}) - h({\bf z}).
  \label{eq:mutual_info}
\end{align}
Since ${\bf z}$ consists of $m$ i.i.d. $\mathcal{N}(0,1)$ random
variables, the entropy of ${\bf z}$ is $0.5\log(2\pi e)^m$.
The entropy of ${\bf y}$ satisfies \cite{Cover1991}
\begin{equation}
  h({\bf y}) \leq \frac{1}{2} \log(2\pi e)^m |{\bf {\bf \Sigma}}_Y|
  \label{}
\end{equation}
with equality achieved if and only if ${\bf y} \sim \mathcal{N}(0,
{\bf \Sigma}_Y)$, where $|\cdot|$ denotes the determinant of a matrix
and ${\bf \Sigma}_Y$ stands for the covariance matrix of ${\bf y}$.
Accordingly, the capacity satisfies
\begin{align}
  C &= \max_{p_{WY} ({\bf w}, {\bf y})} I({\bf w}; {\bf y})
  \leq \max_{p_{Y} ({\bf y})} \frac{1}{2} \log |{\bf \Sigma}_Y|
  \label{ineq:capacity2}
\end{align}
where $p_Y({\bf y})$ denotes the probability function of a random
vector ${\bf y}$.
Therefore, we are interested in the determinant of the covariance
matrix ${\bf \Sigma}_Y$.
According to (\ref{eq:channel}), ${\bf \Sigma}_Y = {\bf \Sigma}_W +
{\bf I}_{m}$ where ${\bf I}_{m}$ is the
$m\times m$ identity matrix.

Throughout the proof, denote the $i$-th element of a vector using
a subscript $i$, e.g., the $i$-th element of ${\bf w}$ is $w_i$.
According to the assumptions (M1) and (M4), we have
\begin{equation}
  \mathbb{E}[w_i] 
  = \mathbb{E}\left[\sum_{j=1}^n \phi(i,j) x_j\right]
  = \sum_{j=1}^n \mathbb{E}[\phi(i,j)] \mathbb{E}[x_j]
  = 0
  \label{}
\end{equation}
for $i \in \{1, 2, \dots, m\}$.
For any $i, j \in \{1, 2, \dots, m\}$, we
have
\begin{align*}
  \mathbb{E}[w_i w_j] 
  &= \mathbb{E} \left[\sum_{p=1}^n \phi(i,p) x_p
  \sum_{q=1}^n \phi(j,q) x_q \right] \\
  &= \sum_{p=1}^n \sum_{q=1}^n \mathbb{E}
  [\phi(i,p) \phi(j,q) x_p x_q].
\end{align*}
According to the assumption (M1), we can further write that
\begin{align*}
  \mathbb{E}[w_i w_j] 
  &= \sum_{p=1}^n \sum_{q=1}^n 
  \mathbb{E} [\phi(i,p)\phi(j,q)]
  \mathbb{E} [x_p x_q].
\end{align*}
Moreover, according to the assumptions (M6) and (M8), for any
$p\neq q$, $\phi(i,p)$ and $\phi(j,q)$ are uncorrelated. 
Thus, the following statement
\begin{align*}
  \mathbb{E} [\phi(i,p)\phi(j,q)] = 0
\end{align*}
is true for $p\neq q$.
Since $\mathbb{E}[x_p^2] = \sigma_X^2$ that follows from assumption (S2), we
obtain
\begin{align}
  \mathbb{E}[w_i w_j] 
  &= \sum_{p=1}^n \mathbb{E} [\phi(i,p)\phi(j,p)] \sigma_X^2.
  \label{eq:cov}
\end{align}

Obviously, depending on the assumption (M9a) or (M9b),
the behavior of ${\bf \Sigma}_W$ differs, and thus, the determinant of
${\bf \Sigma}_Y$ differs.
We discuss the determinant of ${\bf \Sigma}_Y$ in the following
subsections.

\section{Completing Proof of Theorem~\ref{lemma:capacity_corr_1}}
\label{app:proof_alpha<1}
According to (\ref{eq:cov}) and assumptions (M5), (M6), (M8) and
(M9a), we have
\begin{align}
  \mathbb{E} [w_i w_j] &= 
  \sum_{p=1}^n \mathbb{E} [\phi(i,p)\phi(j,p)] \sigma_X^2
  \nonumber \\
  &= \left\{
    \begin{array}{ll}
      \sigma_X^2, & i=j \\
      0, & 1 \leq i,j \leq m_o, i \neq j \\
      0, & m_o+1 \leq i,j \leq m, i \neq j \\
      \frac{\sigma_X^2}{m_o}, & 1 \leq i \leq m_o, m_o+1 \leq j \leq m \\
      \frac{\sigma_X^2}{m_o}, & m_o+1 \leq i \leq m, 1 \leq j \leq
      m_o.
    \end{array}
  \right.
  \label{eq:em1}
\end{align}
Thus, ${\bf \Sigma}_W$ can be divided into four blocks, i.e.,
\begin{align}
  {\bf \Sigma}_W = \left[
    \begin{matrix}
      \sigma_X^2 \cdot {\bf I}_{m_o} & \frac{\sigma_X^2}{m_o}
      \cdot {\bf 1}_{m_o \times m_e} \\
      \frac{\sigma_X^2}{m_o} \cdot {\bf 1}_{m_e \times m_o} &
      \sigma_X^2 \cdot {\bf I}_{m_e}
    \end{matrix}
  \right]
  \label{eq:cov_w_block}
\end{align}
where ${\bf 1}_{m_o \times m_e}$ and ${\bf 1}_{m_e \times m_o}$ are
matrices of all ones of dimension $m_o \times m_e$ and $m_e \times m_o$,
respectively.
Accordingly, ${\bf \Sigma}_Y$ can be written as
\begin{align}
  {\bf \Sigma}_Y 
  &= {\bf \Sigma}_W + {\bf I}_{m} \nonumber \\
  &= \left[
    \begin{matrix}
      (\sigma_X^2 + 1) \cdot {\bf I}_{m_o} 
      & \frac{\sigma_X^2}{m_o}
      \cdot {\bf 1}_{m_o \times m_e} \\
      \frac{\sigma_X^2}{m_o} \cdot {\bf 1}_{m_e \times m_o} 
      & (\sigma_X^2 + 1) \cdot {\bf I}_{m_e}
    \end{matrix}
  \right].
  \label{eq:cov_y_block}
\end{align}
According to Section~9.1.2 of \cite{MatrixCookBook}, the
determinant $|{\bf \Sigma}_Y|$ of the block matrix ${\bf \Sigma}_Y$ can be
calculated as
\begin{align}
  |{\bf \Sigma}_Y| 
    &= (\sigma_X^2 + 1)^m \cdot
    \left[1 - \left(\frac{\sigma_X^2}{\sigma_X^2 + 1}\right)^2
    \cdot \alpha\right].
  \label{eq:det_cov_y}
\end{align}
According to (\ref{eq:SNR}) and (\ref{eq:em1}), the SNR can be
expressed as $\gamma = \sigma_X^2$.
Substituting (\ref{eq:det_cov_y}) into (\ref{ineq:capacity2}), the
upper bound on the capacity can be expressed as a function of $\gamma$
and $\alpha$, i.e.,
\begin{align}
  C &\leq \frac{1}{2} 
  \log \left\{(\gamma + 1)^m \cdot
    \left[1 - \left(\frac{\gamma}{\gamma + 1}\right)^2
    \cdot \alpha\right]\right\} \nonumber \\
    &= \frac{m}{2} \log (\gamma+1)
    + \frac{1}{2} \log \left[1-\left(\frac{\gamma}{\gamma+1}\right)^2 \cdot
    \alpha \right].
    \label{ineq:capacity_corr_1p}
\end{align}
The equality in (\ref{ineq:capacity_corr_1p}) is achieved when ${\bf
y} \sim \mathcal{N}(0, {\bf \Sigma}_Y)$.
This completes the proof. \hfill \IEEEQED

\section{Completing Proof of Theorem~\ref{lemma:capacity_corr_2}}
\label{app:proof_alpha>1}
Recall that ${\bf \Phi}_e$ consists of all rows in $\alpha$ groups
of potential rows constructed as shown in Lemma~\ref{remark:group2}.
In the following, the $m_o$ rows of ${\bf \Phi}_o$ are considered as a group
as well.
Therefore, in ${\bf \Phi}$, we have all rows from $(\alpha+1)$ groups.

First, according to assumption (M5) we have 
\begin{equation}
  \sum_{p=1}^n \mathbb{E} [\phi(i,p)\phi(j,p)] = 1
  \label{eq:corr_1}
\end{equation}
for every $i=j \in \{1, 2, \dots, m\}$.
Second, since any two rows within one of the $(\alpha+1)$ groups
are uncorrelated, we have
\begin{equation}
  \sum_{p=1}^n \mathbb{E} [\phi(i,p)\phi(j,p)] = 0
  \label{eq:corr_2}
\end{equation}
for any pair of
$(i, j) \in \{(i, j) | i \neq j, 1\leq i \leq m, 1\leq j \leq m,
\lceil i/m_o \rceil = \lceil j/m_o \rceil\}$ where $\lceil \cdot
\rceil$ denotes the ceiling function.
Thirdly, as reflected in assumption (M8) and Lemma~\ref{remark:group2}, two
rows taken from two different groups are correlated over one
segment, which indicates that the correlation between the
two rows is $1/m_o$ according to assumptions (M5) and (M7).
Thus, we have
\begin{equation}
  \sum_{p=1}^n \mathbb{E} [\phi(i,p)\phi(j,p)] = 1/m_o
  \label{eq:corr_3}
\end{equation}
for any pair of $(i, j) \in \{(i, j) | \lceil i/m_o \rceil \neq \lceil j/m_o
\rceil, 1\leq i \leq m, 1\leq j \leq m\}$.

Let ${\bf U}(1) = \sigma_X^2\cdot {\bf I}_{m_o}$, and
define ${\bf U}(k+1) \in \mathbb{R}^{(k+1)m_o \times (k+1)m_o}$ for $k=1, 2, \dots,
\alpha$ as follows,
\begin{equation*}
    {\bf U}(k+1) = \left[
    \begin{matrix}
      \sigma_X^2 \cdot {\bf I}_{m_o} & \frac{\sigma_X^2}{m_o}
      \cdot {\bf 1}_{m_o \times km_o} \\
      \frac{\sigma_X^2}{m_o} \cdot {\bf 1}_{km_o \times m_o} &
      {\bf U}(k)
    \end{matrix}
  \right].
\end{equation*}
Combining (\ref{eq:cov}), (\ref{eq:corr_1}), (\ref{eq:corr_2}), and
(\ref{eq:corr_3}), ${\bf \Sigma}_W$ can be written in the following
form
\begin{equation}
    {\bf \Sigma}_W = {\bf U}(\alpha+1) = \left[
    \begin{matrix}
      \sigma_X^2 \cdot {\bf I}_{m_o} & \frac{\sigma_X^2}{m_o}
      \cdot {\bf 1}_{m_o \times m_e} \\
      \frac{\sigma_X^2}{m_o} \cdot {\bf 1}_{m_e \times m_o} &
      {\bf U}(\alpha)
    \end{matrix}
  \right].
  \label{eq:cov_w_block_2}
\end{equation}
Accordingly, ${\bf \Sigma}_Y$ can be written as
\begin{align}
  {\bf \Sigma}_Y 
  &= {\bf \Sigma}_W + {\bf I}_{m} \nonumber \\
  &= \left[
    \begin{matrix}
      (\sigma_X^2 + 1) \cdot {\bf I}_{m_o} 
      & \frac{\sigma_X^2}{m_o}
      \cdot {\bf 1}_{m_o \times m_e} \\
      \frac{\sigma_X^2}{m_o} \cdot {\bf 1}_{m_e \times m_o} 
      & {\bf U}(\alpha) + {\bf I}_{m_e}
    \end{matrix}
  \right].
  \label{eq:cov_y_block_2}
\end{align}
Define a $km_o \times km_o$ matrix 
${\bf V}(k) = 1/(\sigma_X^2+1) \cdot [{\bf U}(k) + {\bf I}_{km_o}]$ 
for $k=1, 2,\dots, \alpha+1$.
Then, ${\bf V}(1) = {\bf I}_{m_o}$, and
\begin{align*}
  {\bf V}(k+1)
  &= 1/(\sigma_X^2+1) \cdot ({\bf U}(k+1) + {\bf I}_{(k+1)m_o})\\
  &= \left[
  \begin{matrix}
    {\bf I}_{m_o} 
    & \frac{\beta}{m_o} \cdot {\bf 1}_{m_o\times km_o} \\
    \frac{\beta}{m_o} \cdot {\bf 1}_{km_o \times m_o} 
    & {\bf V}(k)
  \end{matrix}
  \right]
\end{align*}
where $0 < \beta=\sigma_X^2/(\sigma_X^2+1) < 1$.
Therefore, ${\bf \Sigma}_Y = (\sigma_X^2+1) {\bf V}(\alpha+1)$, and thus
$|{\bf \Sigma}_Y| = (\sigma_X^2+1)^m |{\bf V}(\alpha+1)|$.

Denote the eigenvalues and the corresponding eigenvectors of ${\bf V}(k)$
as $\lambda_i$'s and ${\bf q}_i$'s for $1 \leq i \leq km_o$,
respectively ($k=1,2, \dots. \alpha+1$).
For a pair of $\lambda_i$ and $q_i$, we have
\begin{align}
  ({\bf V}(k) - \lambda_i{\bf I}_{km_o})
  {\bf q}_i = {\bf 0}_{km_o \times 1}
\end{align}
where ${\bf 0}_{km_o \times 1}$ stands for a vector of all zeros with
dimension $km_o \times 1$.
Thus, corresponding ${\bf q}_i$'s comprise the basis of the null
space of
\begin{align}
  &\quad {\bf V}(k) - \lambda_i{\bf I}_{km_o} \nonumber \\
	&=
  \left[
  \begin{array}{ll}
    (1-\lambda_i){\bf I}_{m_o} 
    & \frac{\beta}{m_o} \cdot{\bf 1}_{m_o \times (k-1)m_o} \\
    \frac{\beta}{m_o} \cdot{\bf 1}_{(k-1)m_o \times m_o}
    & {\bf V}(k-1)-\lambda_i {\bf I}_{(k-1)m_o} 
  \end{array}
  \right].
  \label{eq:eig1}
\end{align}
We have the following lemma.
\begin{mylemma}
  The eigenvalues of ${\bf V}(k)$ have three different values.
  \begin{itemize}
    \item $\lambda_i = 1$ and there are $(m_o-1)k$ corresponding
      eigenvectors ${\bf q}_i$. 
      They satisfy ${\bf 1}_{1\times km_o} {\bf q}_i = 0$.
    \item $\lambda_i = 1-\beta$ and there are $(k-1)$ corresponding
      eigenvectors ${\bf q}_i$. 
      They satisfy ${\bf 1}_{1\times km_o} {\bf q}_i = 0$.
    \item $\lambda_i = 1+(k-1)\beta$ and there is a
      single corresponding eigenvector ${\bf q}_i$. 
      It satisfies ${\bf 1}_{1\times km_o} {\bf q}_i = \sqrt{km_o}$.
  \end{itemize}
\end{mylemma}
\begin{IEEEproof}
  Divide the matrix shown in (\ref{eq:eig1}) into $k$ sub-matrices,
  with the $i$-th ($i=1,2, \dots, k$) sub-matrix ${\bf B}_i \in \mathbb{R}^{km_o \times m_o}$
  consisting of the $[(i-1)m_o+1]$-th column to the $im_o$-th column.

  When $\lambda_i=1$, the diagonal elements of the
  matrix in (\ref{eq:eig1}) are all zeros.
  It can be observed that within each sub-matrix ${\bf B}_i$, the columns are
  identical.
  Since ${\bf q}_i$'s comprise the basis of the null space of
  (\ref{eq:eig1}), there are $(m_o-1)k$ such eigenvectors and
  ${\bf 1}_{1\times km_o} {\bf q}_i = 0$.

  When $\lambda_i=1-\beta$, the diagonal elements of
  the matrix in (\ref{eq:eig1}) are all equal to $\beta$.
  It can be observed that for each sub-matrix ${\bf B}_i$, we have
  ${\bf B}_i {\bf 1}_{m_o\times 1} = \beta {\bf 1}_{km_o\times 1}$.
  Thus, there are $(k-1)$ corresponding eigenvectors and ${\bf 1}_{1\times
  km_o} {\bf q}_i = 0$.

  When $\lambda_i=1+(k-1)\beta$, the diagonal elements of the matrix in
  (\ref{eq:eig1}) are all equal to $-(k-1)\beta$.
  It can be observed that 
  $[{\bf V}(k) - \lambda_i{\bf I}_{km_o}] {\bf 1}_{km_o \times 1}$ is a vector
  of zeros.
  Thus, there is a single corresponding eigenvector and it satisfies  
  ${\bf 1}_{1\times km_o} {\bf q}_i = \sqrt{km_o}$
  considering that $||{\bf q}_i||_2 = 1$.

  The matrix ${\bf V}(k)$ is symmetric, and thus it has totally
  $km_o$ mutually orthogonal eigenvectors.
  We have already found all of them. 
  Thus, there are no other eigenvalues and eigenvectors.
  This completes the proof.
\end{IEEEproof}

Using these remarks, the determinant of ${\bf V}(k)$ can be obtained as
\begin{align}
  |{\bf V}(k)| = (1-\beta)^{k-1} (1+(k-1)\beta).
  \label{eq:det_V}
\end{align}
Accordingly, 
\begin{align*}
  |{\bf \Sigma}_Y| & = (\sigma_X^2 + 1)^m |{\bf V}(\alpha+1)| \\
  &= (\sigma_X^2 + 1)^m (1-\beta)^\alpha (1+\alpha \beta).
\end{align*}
Noting that $\gamma = \sigma_X^2$ and $\beta =
\sigma_X^2/(\sigma_X^2+1)$,
the upper bound on the capacity $C$ in (\ref{ineq:capacity2}) can be
expressed as
\begin{align}
  C &\leq \frac{1}{2} 
  \log \left[(\gamma + 1)^m \cdot
  \left(1- \frac{\gamma}{\gamma+1}\right)^\alpha 
  \cdot \left(1+\frac{\alpha\gamma}{\gamma+1}\right) \right] \nonumber \\
  &= \frac{m}{2} \log (\gamma\!+\!1)
   \!-\!\frac{\alpha+1}{2} \log (\gamma\!+\!1)
   \!+\!\frac{1}{2} \log \left[ (1\!+\!\alpha)\gamma\!+\!1 \right].
   \label{ineq:capacity_corr_2p}
\end{align}
The equality in (\ref{ineq:capacity_corr_2p}) is achieved when ${\bf
y} \sim \mathcal{N}(0, {\bf \Sigma}_Y)$.
This completes the proof. \hfill \IEEEQED



\begin{thebibliography}{10}

\bibitem{Candes2005a}
E.~J.~Cand\`{e}s and T.~Tao, ``Decoding by linear programming,'' \emph{{IEEE}
  Trans. Inf. Theory}, vol.~51, no.~12, pp.~4203--4215, Dec.~2005.

\bibitem{Candes2006c}
E.~J.~Cand\`{e}s, ``Compressive sampling,'' in \emph{Proc. Int. Cong. Math.},
  Madrid, Spain, Aug.~2006, pp.~1433--1452.

\bibitem{Candes2008}
E.~J.~Cand\`{e}s and M.~B.~Wakin, ``An introduction to compressive sampling,''
  \emph{{IEEE} Signal Process. Mag.}, vol.~25, no.~2, pp.~21--30, Mar.~2008.

\bibitem{Donoho2006}
D.~L.~Donoho, ``Compressed sensing,'' \emph{{IEEE} Trans. Inf. Theory},
  vol.~52, no.~4, pp.~1289--1306, Apr.~2006.

\bibitem{Laska2007}
J.~N.~Laska, S.~Kirolos, M.~F.~Duarte, T.~S.~Ragheb, R.~G.~Baraniuk, and
  Y.~Massoud, ``Theory and implementation of an analog-to-information converter
  using random demodulation,'' in \emph{Proc. IEEE Int. Symp. Circuits and
  Systems}, New Orleans, LA, May~2007, pp.~1959--1962.
	
\bibitem{Becker}
S.~R.~Becker, \emph{Practical Compressed Sensing: Modern Data Acquisition and
Signal Processing}.\hskip 1em plus 0.5em minus 0.4em\relax PhD Thesis: California 
Institute of Technology, Pasadena, California, 2011.

\bibitem{Taheri2010}
O.~Taheri and S.~A.~Vorobyov, ``Segmented compressed sampling for
  analog-to-information conversion,'' in \emph{Proc. Computational Advances in
  Multi-Sensor Adaptive Process. (CAMSAP)}, Aruba, Dutch Antilles, Dec.~2009, 
	pp.~113--116.

\bibitem{Taheri2011}
O.~Taheri and S.~A.~Vorobyov, ``Segmented compressed sampling for analog-to-information 
conversion: Method and performance analysis,'' \emph{{IEEE} Trans. Signal Process.},
  vol.~59, no.~2, pp.~554--572, Feb.~2011.

\bibitem{Wainwright2009}
M.~Wainwright, ``Information-theoretic limits on sparsity recovery in the
  high-dimensional and noisy setting,'' \emph{{IEEE} Trans. Inf. Theory},
  vol.~55, no.~12, pp.~5728--5741, Dec.~2009.

\bibitem{Wang2010a}
W.~Wang, M.~J.~Wainwright, and K.~Ramchandran, ``Information-theoretic limits
  on sparse signal recovery: Dense versus sparse measurement matrices,''
  \emph{{IEEE} Trans. Inf. Theory}, vol.~56, no.~6, pp.~2967--2979, Jun.~2010.

\bibitem{Reeves2008}
G.~Reeves and M.~Gastpar, ``Sampling bounds for sparse support recovery in the
  presence of noise,'' in \emph{Proc. Int. Symp. Inf. Theory}, Toronto, Canada,
  Jul.~2008, pp.~2187--2191.

\bibitem{Akcakaya2010}
M.~Akcakaya and V.~Tarokh, ``Shannon-theoretic limits on noisy compressive
  sampling,'' \emph{{IEEE} Trans. Inf. Theory}, vol.~56, no.~1, pp.~492--504,
  Jan.~2010.

\bibitem{Aeron2010}
S.~Aeron, V.~Saligrama, and M.~Zhao, ``Information theoretic bounds for
  compressed sensing,'' \emph{{IEEE} Trans. Inf. Theory}, vol.~56, no.~10, 
	pp.~5111--5130, Oct.~2010.

\bibitem{Wainwright2009a}
M.~J.~Wainwright, ``Sharp threshold for high-dimensional and noisy sparsity
  recovery using $\ell_1$-constrained quadratic programming ({Lasso}),''
  \emph{{IEEE} Trans. Inf. Theory}, vol.~55, no.~5, pp.~2183--2202, May~2009.

\bibitem{Reeves2013}
G.~Reeves and M.~Gastpar, ``Approximate sparsity pattern recovery:
  Information-theoretic lower bounds,'' \emph{{IEEE} Trans. Inf. Theory},
  vol.~59, no.~6, pp.~3451--3465, Jun.~2013.

\bibitem{Baraniuk2008}
R.~G.~Baraniuk, M.~Davenport, R.~DeVore, and M.~Wakin, ``A simple proof of the
  restricted isometry property for random matrices,'' \emph{Constructive
  Approximation}, vol.~28, no.~3, pp.~253--263, Dec.~2008.
	
	\bibitem{Haupt2006}
J.~Haupt and R.~Nowak, ``Signal reconstruction from noisy random projections,''
  \emph{{IEEE} Trans. Inf. Theory}, vol.~52, no.~9, pp.~4036--4048, Sep.~2006.

\bibitem{Fletcher2007}
A.~K.~Fletcher, S.~Rangan, and V.~Goyal, ``On the rate-distortion performance
  of compressed sensing,'' in \emph{Proc. IEEE Int. Conf. Acoustics, Speech and
  Signal Process.}, Honolulu, HI, Apr.~2007, pp.~885--888.

\bibitem{Sarvotham2006}
S.~Sarvotham, D.~Baron, and R.~G.~Baraniuk, ``Measurement vs. bits: Compressed
  sensing meets information theory,'' in \emph{Proc. Allerton Conf.
  Communication, Control and Computing}, Allerton, IL, USA, Sep.~2006,
  pp.~1419--1423.

\bibitem{Candes2010}
E.~J.~Cand\`{e}s, Y.~C.~Eldar, D.~Needell, and P.~Randall, 
``Compressed sensing with coherent and redundant dictionaries,'' 
\emph{Applied and Computational Harmonic Analysis}, vol.~531, no.~1, 
pp.~59--73, Jan.~2011.

\bibitem{TaheriAsilomar}
O.~Taheri and S.~A.~Vorobyov, ``Empirical risk minimization-based analysis of segmented 
compressed sampling,'' in \emph{Proc. 44th Annual Asilomar Conf. Signals, Systems, and 
Computers}, Pacific Grove, California, USA, Nov.~2010, pp.~233--235.

\bibitem{LinearCongEqMathworld}
E.~W.~Weisstein, ``Linear congruence equation,'' in \emph{MathWorld},
  \url{http://mathworld.wolfram.com/LinearCongruenceEquation.html}.

\bibitem{Gamal2012}
A.~E.~Gammal and Y.-H.~Kim, \emph{Network Information Theory}.\hskip 1em plus
  0.5em minus 0.4em\relax New York: Cambridge Univ. Press, 2012.

\bibitem{Berger1971}
T.~Berger, \emph{Rate Distortion Theory: A Mathematical Basis for Data
  Compression}.\hskip 1em plus 0.5em minus 0.4em\relax Englewood Cliffs, NJ:
  Prentice-Hall, 1971.

\bibitem{Cover1991}
T.~M. Cover and J.~A. Thomas, \emph{Elements of Information Theory}.\hskip 1em
  plus 0.5em minus 0.4em\relax New York: Wiley, 2011.

\bibitem{MatrixCookBook}
K.~B. Petersen and M.~S. Pedersen, \emph{The Matrix Cookbook}, 2008.

\end{thebibliography}
\end{document}